\documentclass[12pt,thmsa]{article}
\usepackage{sw20lart}

\input tcilatex
\QQQ{Language}{
American English
}

\begin{document}

\author{I. Radinschi\thanks{%
jessica@etc.tuiasi.ro} \and ''Gh. Asachi'' Technical University, Iasi,
Romania}
\title{Energy Distribution of a Schwarzschild Black Hole in a Magnetic Universe }
\maketitle

\begin{abstract}
We obtain the energy distribution of a Schwarzschild black hole in a
magnetic universe in the Tolman prescription.

PACS: 04. 20.-q; 04. 70.-s

\end{abstract}

\section{INTRODUCTION}

The localization of energy is a long-standing problem in the theory of
general relativity. Numerous attempts have been made for a solution.

Virbhadra and his collaborators investigated the problem of the
energy-momentum localization by using the energy-momentum complexes. The
results obtained for several particular space-times (the Kerr-Newman, the
Einstein-Rosen and the Bonnor-Vaidya) lead to the conclusion that different
energy-momentum complexes give the same energy distribution for a given
space-time [1]-[6]. Aguirregabiria, Chamorro and Virbhadra [7] showed that
several energy-momentum complexes coincide for any Kerr-Schild class metric.
Xulu obtained interesting results about the energy distribution of a charged
dilaton black hole [8] and about the energy associated with a Schwarzschild
black hole in a magnetic universe [9]. Also, recently, Xulu [10] obtained
the total energy of a model of universe based on the Bianchi I type metric.
The author calculated the energy distribution of a dyonic dilaton black hole
[11] and the energy of the Bianchi type I solution [12]. Also, we obtained
the energy distribution in a static spherically symmetric nonsingular black
hole space-time [13]. Recently, Virbhadra [14] shows that different
energy-momentum complexes give the same and reasonable results for many
space-times.

Xulu [9] obtained the energy associated with a Schwarzschild black hole in a
magnetic universe. Melvin's magnetic universe [15] is a solution of the
Einstein-Maxwell equations corresponding to a collection of parallel
magnetic lines of force held together by mutual gravitation. The physical
structure of the magnetic universe and its dynamical behavior was studied by
Thorne. Ernst [16] obtained axially symmetric exact solution to the
Einstein-Maxwell equations representing a Schwarzschild black hole immersed
in Melvin's uniform magnetic universe.

The purpose of this paper is to compute the energy distribution for the
Ernst space-time by using the Tolman prescription. We use the geometrized
units $(G=1,c=1)$ and follow the convention that the Latin indices run from $%
0$ to $3$.

\section{\ ENERGY\ IN\ TOLMAN'S\ PRESCRIPTION}

It is interesting to evaluate the energy distribution of a magnetic black
hole. We know that the Einstein-Maxwell equations are

\begin{equation}
R_i^{\,\;k}-\frac 12g_i^{\;k}R=8\pi T_i^{\;k},
\end{equation}

\begin{equation}
F_{ij,k}+F_{jk,i}+F_{ki,j}=0,
\end{equation}

\begin{equation}
\frac 1{\sqrt{-g}}(\sqrt{-g}F^{ik})_{,k}=4\pi J^i.
\end{equation}

The energy-momentum of the electromagnetic field is given by

\begin{equation}
T_i^{\;k}=\frac 1{4\pi }(-F_{im}F^{km}+\frac 14g_i^{\;k}F_{mn}F^{mn}).
\end{equation}

Ernst [16] obtained the axially symmetric electrovac solution $(J^i=0)$ to
the equations (1), (2) and (3) that corresponds to a Schwarzschild black
hole in Melvin's magnetic universe.

The metric is given by

\begin{equation}
ds^2=\Delta ^2[(1-\frac{2M}r)dt^2-(1-\frac{2M}r)^{-1}dr^2-r^2d\theta
^2]-\Delta ^{-2}r^2\sin ^2\theta d\varphi ^2.
\end{equation}

The Cartan components of the magnetic field are given by

\begin{equation}
\begin{tabular}{c}
$H_r=\Delta ^{-2}B_0\cos \theta ,$ \\ 
$H_\theta =-\Delta ^{-2}B_0(1-\frac{2M}r)^{\frac 12}\sin \theta ,$%
\end{tabular}
\end{equation}

where

\begin{equation}
\Delta =1+\frac 14B_0^2r^2\sin ^2\theta
\end{equation}

and $M$ and $B_0$ are constants. We note that the Ernst solution is a black
hole solution and $r=2M$ is the event horizon.

The Tolman energy-momentum complex [17] is given by

\begin{equation}
\Upsilon _i^{\;k}=\frac 1{8\pi }U_{i\;\;\;,}^{\;kl}{}_l,
\end{equation}

where $\Upsilon _0^{\;0}$ and $\Upsilon _\alpha ^{\;\,0}$ are the energy and
momentum components.

\begin{equation}
U_i^{\;kl}=\sqrt{-g}(-g^{pk}V_{ip}^{\;\;l}+\frac
12g_i^{\;k}g^{pm}V_{pm}^{\;\;\;\,l}),
\end{equation}

with

\begin{equation}
V_{jk}^{\;\;i}=-\Gamma _{jk}^i+\frac 12g_j^{\;i}\Gamma _{mk}^m+\frac
12g_k^{\;i}\Gamma _{mj}^m.
\end{equation}

Also, the energy-momentum complex $\Upsilon _i^{\;k}$ also satisfies the
local conservation laws

\begin{equation}
\frac{\partial \Upsilon _i^{\;k}}{\partial x^k}=0.
\end{equation}

The energy and momentum in Tolman prescription are given by

\begin{equation}
P_i=\iiint \Upsilon _i^{\;0}dx^1dx^2dx^3.
\end{equation}

Using Gauss's theorem we obtain

\begin{equation}
P_i=\frac 1{8\pi }\iint U_i^{\;0\alpha }n_\alpha dS,
\end{equation}

where $n_\alpha =(\frac xr,\frac yr,\frac zr)$ are the components of a
normal vector over an infinitesimal surface element $dS=r^2\sin \theta
d\theta d\varphi $.

The Tolman energy-momentum complex gives the correct result if the
calculations are carried out in quasi-Cartesian coordinates. We transform
the line element (1) to quasi-Cartesian coordinates $t,x,y,z$ according to

\begin{equation}
\begin{tabular}{c}
$r=(x^2+y^2+z^2)^{\frac 12},$ \\ 
$\theta =\cos ^{-1}(\frac z{\sqrt{x^2+y^2+z^2}}),$ \\ 
$\varphi =\tan ^{-1}(y/x).$%
\end{tabular}
\end{equation}

The metric (1) becomes [9]

\begin{eqnarray}
ds^2 &=&\Delta ^2(1-\frac{2M}r)dt^2-[\Delta ^2(\frac{ax^2}{r^2})+\Delta
^{-2}(\frac{y^2}{x^2+y^2})]dx^2-[\Delta ^2(\frac{ay^2}{r^2})+ \\
&&+\Delta ^{-2}(\frac{x^2}{x^2+y^2})]dy^2-\Delta ^2[1+\frac{2Mz^2}{r^2(r-2M)}%
]dz^2-[\Delta ^2(\frac{2axy}{r^2})+\Delta ^{-2}  \nonumber \\
&&(-\frac{2xy}{x^2+y^2})]dxdy-\Delta ^2[\frac{4Mxz}{r^2(r-2M)}]dxdz-\Delta
^2[\frac{4Myz}{r^2(r-2M)}]dydz,  \nonumber
\end{eqnarray}

where

\begin{equation}
a=\frac{2M}{r-2M}+\frac{r^2}{x^2+y^2}.
\end{equation}

The components of the Tolman energy-momentum complex are calculated with the
Maple GR Tensor II Release 1.50.

Because the components of the $U_i^{\;kl}$ are too many we give only those
which are involved in the calculation of the energy

\begin{equation}
\begin{tabular}{c}
$U_0^{\;01}=\frac{2Mx}{r^3}+\frac{(\Delta ^4-1)}2[\frac x{x^2+y^2}],$ \\ 
$U_0^{\;02}=\frac{2My}{r^3}+\frac{(\Delta ^4-1)}2[\frac y{x^2+y^2}],$ \\ 
$U_0^{\;03}=\frac{2Mz}{r^3}.$%
\end{tabular}
\end{equation}

Using (12), (17) and applying (13) we obtain the energy distribution for the
Ernst space-time 
\begin{equation}
E(r)=M+\frac 1{8\pi }\int_{\theta =0}^0\int_{\varphi =0}^{2\pi }\frac{%
(\Delta ^4-1)}2r\sin \theta d\theta d\varphi .
\end{equation}

We replace in (18) the value of $\Delta $ from (7) and consider the values
of $G$ and $c$ and we have

\begin{equation}
E(r)=Mc^2+\frac 16B_0^2r^3+\frac 1{20}\frac G{c^4}B_0^4r^5+\frac 1{140}\frac{%
G^2}{c^8}B_0^6r^7+\frac 1{2520}\frac{G^3}{c^{12}}B_0^8r^9.
\end{equation}

The relation (19) can be also written

\begin{equation}
E(r)=Mc^2+\frac 1{8\pi }\iiint B_0^2dV+\frac 1{20}\frac G{c^4}B_0^4r^5+\frac
1{140}\frac{G^2}{c^8}B_0^6r^7+\frac 1{2520}\frac{G^3}{c^{12}}B_0^8r^9.
\end{equation}

In the expression of the energy distribution the first term represents the
rest mass-energy of the Schwarzschild black hole, the second is the special
relativistic value for the energy of the uniform magnetic field and the
other terms that remain in the expression are due to the general
relativistic effect.

\section{DISCUSSION}

The main purpose of the present paper is to show that the problem of the
localization of energy in relativity can be solved by using the
energy-momentum complexes. The Bondi opinion [18] is that a nonlocalizable
form of energy is not admissible in relativity so its location can in
principle be found. Some interesting results which have been found recently
[7]-[13], support the idea that the several energy-momentum complexes can
give the same and acceptable result for a given space-time. Also, in his
recent paper Virbhadra [14] emphasized that though the energy-momentum
complexes are non-tensors under general coordinate transformations, the
local conservation laws with them hold in all coordinate systems. Chang,
Nester and Chen [19] showed that the energy-momentum complexes are actually
quasilocal and legitimate expressions for the energy-momentum.

We calculated the energy distribution for the Ernst space-time in the Tolman
prescription and find the same result as the result obtained by S. S. Xulu
[9] in the Einstein prescription. The energy increases because of the
presence of the magnetic field.

\end{document}